\documentclass[usenatbib]{mn2e}
\usepackage{graphicx}

\title[GOES12 Remote Sounding]{Fraction of clear skies above astronomical sites: a new analysis from the GOES12 satellite}

\author[S. Cavazzani et al.]{S. Cavazzani$^{1}$\thanks{E-mail:stefano.cavazzani@unipd.it}, S. Ortolani$^{1}$,
 V.Zitelli $^{2}$, Y.Maruccia$^{3}$\\
$^{1}$Department of Astronomy, University of Padova, Vicolo
dell'Osservatorio 3, I-35122, Padova, Italy\\
$^{2}$INAF-Osservatorio Astronomico di Bologna, via Ranzani 1, I-40127, Bologna, Italy\\
$^{3}$Department of Physics, University of Salento, Via per Arnesano, CP 193, 73100 Lecce, Italy}

\begin{document}

\date{Accepted 2010 September 24.  Received 2010 September 24; in original form 2010 May 25.}

\pagerange{\pageref{firstpage}--\pageref{lastpage}} \pubyear{2009}

\maketitle

\label{firstpage}

\begin{abstract}

Comparing the number of clear nights (cloud free) available for astronomical observations is a critical task because it should be based on homogeneous methodologies. Current data are mainly based on different judgements based on observer logbooks or on different instruments.
In this paper we present a new homogeneous methodology on very different astronomical sites for modern optical astronomy, in order to quantify the available night time fraction. The data are extracted from night time GOES12 satellite infrared images and compared with ground based conditions when available.  In this analysis we introduce a wider average matrix and 3-Bands correlation in order to reduce the noise and to distinguish between clear and stable nights. Temporal data are used for the classification.
In the time interval 2007-2008 we found that the percentage of the satellite clear nights is $88\%$ at Paranal, $76\%$ at La Silla, $72.5\%$ at La Palma, $59\%$ at Mt. Graham and $86.5\%$ at Tolonchar.
The correlation analysis of the three GOES12 infrared bands B3, B4 and B6 indicates that the fraction of the stable nights is lower by $2\%$ to $20\%$ depending on the site.

\end{abstract}


\begin{figure*}
  \centering
  \includegraphics[width=14cm]{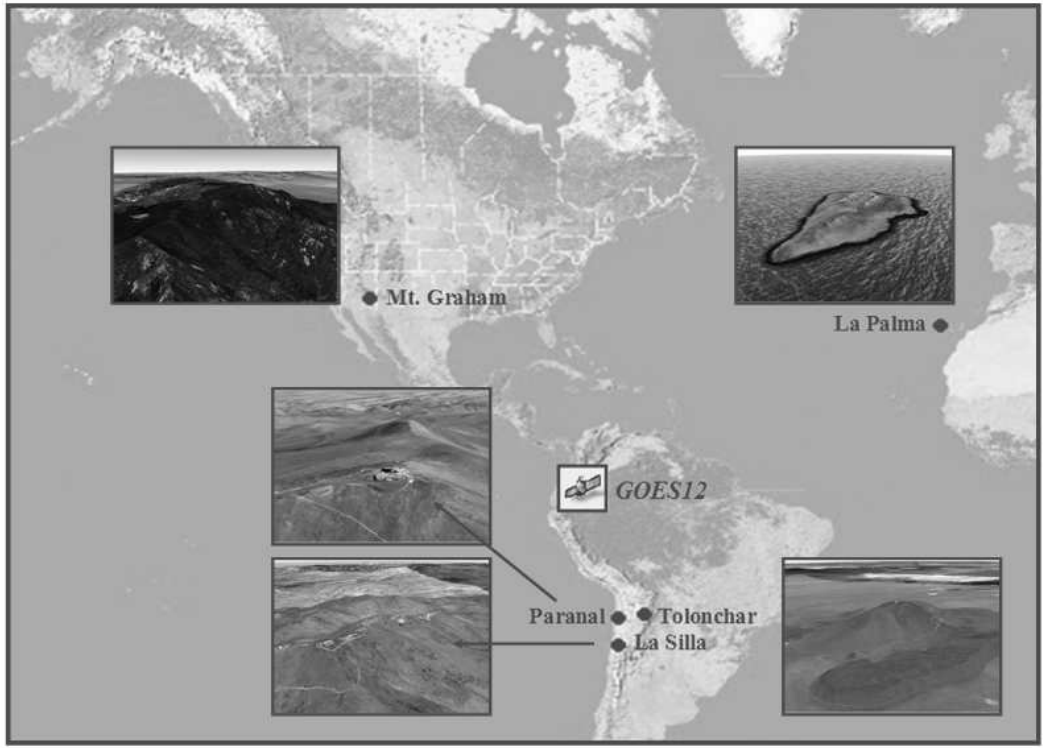}
  \caption{Location of the five sites involved in the analysis. As seen in the inserts the selected sites presents very different topographic
conditions: La Palma is a sharp island, Mt. Graham a relatively wide
plateau, Paranal, Tolonchar and La Silla are isolated peaks over a
desertic altopiano. The position of GOES12 satellite projected on the map.}
             \label{m}
\end{figure*}

\section{Introduction}

The efficiency of the astronomical telescopes is critically
dependent on the cloud coverage. The knowledge of the clear night 
time fraction is then fundamental for the choice of a telescope
site, and, on already existing facilities, its distribution during
the year, as well as long term trends, are very important for
planning the observations and the development  of the
instrumentation. In the last century the quantification of the night
time clear fraction was based mainly on specific visual inspection
of the sky conditions or on the observational logbooks of the
telescopes.  These methods are "internally" robust, but they are
dependent on the experience of the observer and on the quality of
the site. In short time tests there could be also some dependence on
the Moon phase. An adequate time coverage is time consuming and
expensive when applied to several, new sites. The use of the
archives of satellite images allows, instead, to investigate
simultaneously several sites in a time base of several years. In
this study night time satellite derived parameters are used to
assess the clear-usable fraction from the Geostationary Operational
Environmental Satellite 12 (GOES12) archive. The selected sites  are
chosen in order to test different climatic conditions and are located in Chile, USA and Spain (Canary Islands), (See Fig. \ref{m}).\\
Most of them are already well known, developed sites and host large,
modern telescopes. The site of Tolonchar, instead, have been studied
during the Thirty Meter Telescope (TMT) survey, but it is located in relatively little known
area for the optical near-infrared astronomy.  In this analysis we have a double goal:
to check the reliability of our analysis method, to explore the
characteristics of new regions and to compare them.
The GOES satellite data have been studied with the goal to study environmental conditions, but they have
been recently used also for the study of cloud coverage and water
vapor content above some astronomical sites (Erasmus \& van Rooyen (\cite{erasmus06}), della Valle et al. (\cite{dellav10}) (PaperIII)). The advantage of GOES over other satellites is to have a very stable and very high orbit, allowing 
the collection of simultaneous images of almost half of the Earth hemisphere, still with a high
resolution (4 km in the infrared (IR)). In this way site to site random
biases, due to instrument instabilities are reduced. Furthermore the
infrared channels allow the detection of the thermal radiation
emitted during the night from different atmospheric layers and/or
from the soil. An appropriate choice of the wavelength allows to
choose the optimal layer emission height above the site. If it
occurs well above the soil surface, the signal becomes
independent of the specific soil properties and of low level
conditions. Phenomena occurring below the selected site (fog, low
clouds...) are also avoided. In some sites, for example at La Palma,
this aspect is of crucial importance. The channels used in our
analysis have been selected with the above discussed criteria and
are summarized in Table \ref{BAND}. In a previous paper (Paper III) we studied the clear
sky fraction at La Palma and Mt.Graham, from ground and satellite,
using an approach similar to Erasmus \& van Rooyen (\cite{erasmus06}), but we have used direct GOES12 satellite brightness temperature measurements. 
In this paper the descriptions of the adopted definitions used to classify the nights are reported in Sections \ref{cmcnc} and \ref{csnc}.\\
The new adopted method is validated using the database of La Palma and Paranal and, after the positive results, it is applied to the other sites under investigation.
The paper is organized as follows: 

\begin{itemize}
	\item in Section \ref{tud} we describe the used database,
	\item in Section \ref{satacqui} we describe the satellite data acquisition procedure,
	\item in Section \ref{rsbm} we describe the mathematical used model,
	\item in Section \ref{analysis} we describe the IR analysis,
	\item in Section \ref{acf} we describe the atmospheric correlation function,
	\item in Section \ref{hda} we report the data analysis and discussion of the results.
\end{itemize}

\begin{table}
 \centering
 \begin{minipage}{80mm}
  \caption{GOES12 bands and resolution at Nadir.}
   \label{BAND}
  \begin{tabular}{@{}lcccc@{}}
  \hline
                & Window  & Passband     & Resolution      \\
                &         & $[\mu m]$     &          [km]           \\
 \hline
 \textit{BAND1} &Visible  &    $0,55\div0.75$ &  $4$\\
 \textit{BAND2} &   Microwaves  &     $3.80\div4,00$ & $4$  \\
 \textit{BAND3} &$H_{2}O$   &   $6,50\div7.00$ &  $4$\\
	\textit{BAND4}   & $IR$   & $10,20\div11.20$ & $4$\\
	\textit{BAND6}&   $CO_{2}$  &    $13.30$ &  $8$\\
 \hline

\end{tabular}
\end{minipage}
\end{table}

\section{The used database}
\label{tud}

In this analysis we have used several sets of data collected from ground and satellite facilities partially
available via web and partially obtained thanks to the courtesy of the observatory staff.
The validation of satellite data are also performed via correlations among ground based and satellite data. 
In this paper we have sampled the years 2007 and 2008. Table \ref{datacomp} shows the characteristics of the used databases.

\subsection{Ground Based Data}
\label{gbd}

Differences at La Palma microclimate have been discussed in previous papers (Lombardi et al. (\cite{lombardi06}) (hereafter Paper I), Lombardi et al. (\cite{lombardi07}) (hereafter Paper II), and della Valle et al.(\cite{dellav10}) (Paper III)). Paper I shows a complete analysis of the vertical temperature gradients and their correlation with the astronomical seeing, Paper II shows an analysis of the correlation between wind
and astronomical parameters as well as the overall long term weather conditions at La Palma. A statistical fraction of clear nights from satellite has been derived in Paper III using a basic approach to test the ability of the satellite to select clear nights.
In order to have a reference for a classification of the nights at La
Palma we used three different sources: the logbooks obtained from TNG (Telescopio Nazionale Galileo) and
from the Liverpool telescope, and the data from the TNG meteorological
station. The logbooks have been used merging the information, filling the
gaps and checking the comments in case of contradictory classifications.
The data from TNG meteorological station have been used to understand the
status of ambiguous or unclassified nights in terms of humidity or wind
speed limits.
In general the agreement was good, but in winter time all the three
sources were often needed in order to have a realistic view of the night
weather evolution.
The study of the telescope logbooks at Paranal was not needed because  the
night status data are obtained from the web pages of the ESO (European Southern Observatory) Observatories Ambient Conditions Database\footnote{See http://archive.eso.org/asm/ambient-server}. They are very detailed pages containing the hourly humidity, temperature, atmospheric pressure, direction and wind speed. In addition there are measures of seeing through the DIMM (Differential Image Motion Monitor) and measures of the flux of a reference star. In particular the flux can trace the presence of clouds.
The same website is also available for La Silla, but unfortunately in this case the web page is less detailed and often the data are missing.
La Silla database is used as a further check.
Both the sites of La Palma and Paranal are used to test and validate the new model applied in this paper.

\subsection{Satellite Based Data}
\label{goes12}

In these last decades the site testing have been conducted adding to the traditional meteorological instruments the use of the satellite data. Satellite archives contain several parameters useful for astronomical observations, allowing to compare different sites in a suitable way. Varela et al. (\cite{varela}) give an exhaustive presentation of the satellites used for site testing. In our analysis we have chosen among the other available satellites the GOES satellite because it detects the IR night time emitted radiation. A detailed discussion is presented in Sec. \ref{advantage}.
GOES is an American geosynchronous weather facilities of the National Oceanic and Atmospheric Administration (NOAA), and it is able to observe the full Earth disk. It is designed to detect surface temperature and the cloud cover, in addition to other important meteorological parameters. GOES12 have on board an imager covering five wavelength channels, one in the visible bands and four in the infrared bands (see Table \ref{BAND}). The maximum temporal resolution of the full Earth-disk scans is $41~sec$ that is a very high temporal sampling. Moreover GOES have also a high spatial resolution. It should be noticed that GOES12 observed La Palma area at $64^{\circ}10'$ from Nadir, near the edge of the field of view (Table \ref{ST}).

\begin{table}
 \centering
 \begin{minipage}{80mm}
  \caption{Total amount of consecutive nights covered by each databases.}
   \label{datacomp}
  \begin{tabular}{@{}lccc@{}}
  \hline

 Site       & Ground Data & Satellite Data &  \\

 \hline
 Paranal   & 730 & 700    &  \\
 La Silla  & 730 & 700    &  \\
 La Palma  & 730 & 700  &  \\
 Mt.Graham &     & 700 &  \\
 Tolonchar &     & 700 &  \\
 \hline

\end{tabular}
\end{minipage}
\end{table}

\subsubsection{Advantages of GOES12 Satellite}
\label{advantage}

We preferred to use GOES among the other satellite for several reasons that are explained below:

\begin{itemize}
	\item Because it is possible to observe, with a single image, several sites simultaneously.
	\item Because, thanks to the very high orbit ($35800km$), the satellite is extremely stable and not affected by phenomena of high exosphere.
	\item Because, thanks to this set-up, it is possible to have the same instrumental configuration for each site and to compare them in a suitable way.
	\item Because GOES12 data have a high temporal resolution (41 sec as maximum value) and the complete day coverage.
    \item Because GOES12 observe the site at any time of the day. Instead polar satellites are bound to individual orbits. This allows to use an hourly analysis instead of a daily average of atmospheric conditions.
   \item Because GOES12 data have a high spatial resolution (1 km for visual to 4 km in IR bands).
	\item Because GOES12 provides five simultaneous images, one for each band, and it is the only satellite with the $CO_{2}$ band ($13,30\mu m$) very useful for the analysis of lower atmosphere phenomena.
    \item Because GOES12 have a long term database, useful for long time analysis.
    \item Because presents the same deterioration of images due to the inevitable degrade of the satellite. As a consequence the comparison between different sites is not influenced by the use of different instruments or different images.
\end{itemize}

\begin{figure}
  \centering
   \includegraphics[width=8.5cm]{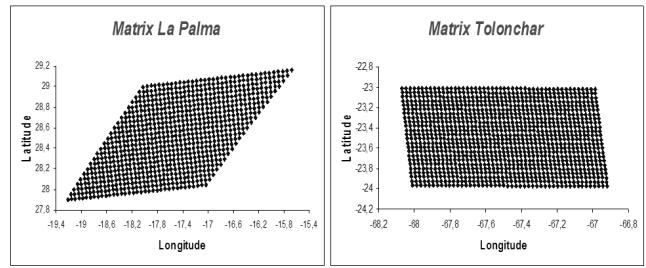}
  \caption{Comparison of one image matrix at La Palma and Tolonchar. The deformation is due to the satellite observation angle.}
             \label{cmat}
   \end{figure}

\begin{figure}
  \centering
  \includegraphics[width=8.5cm]{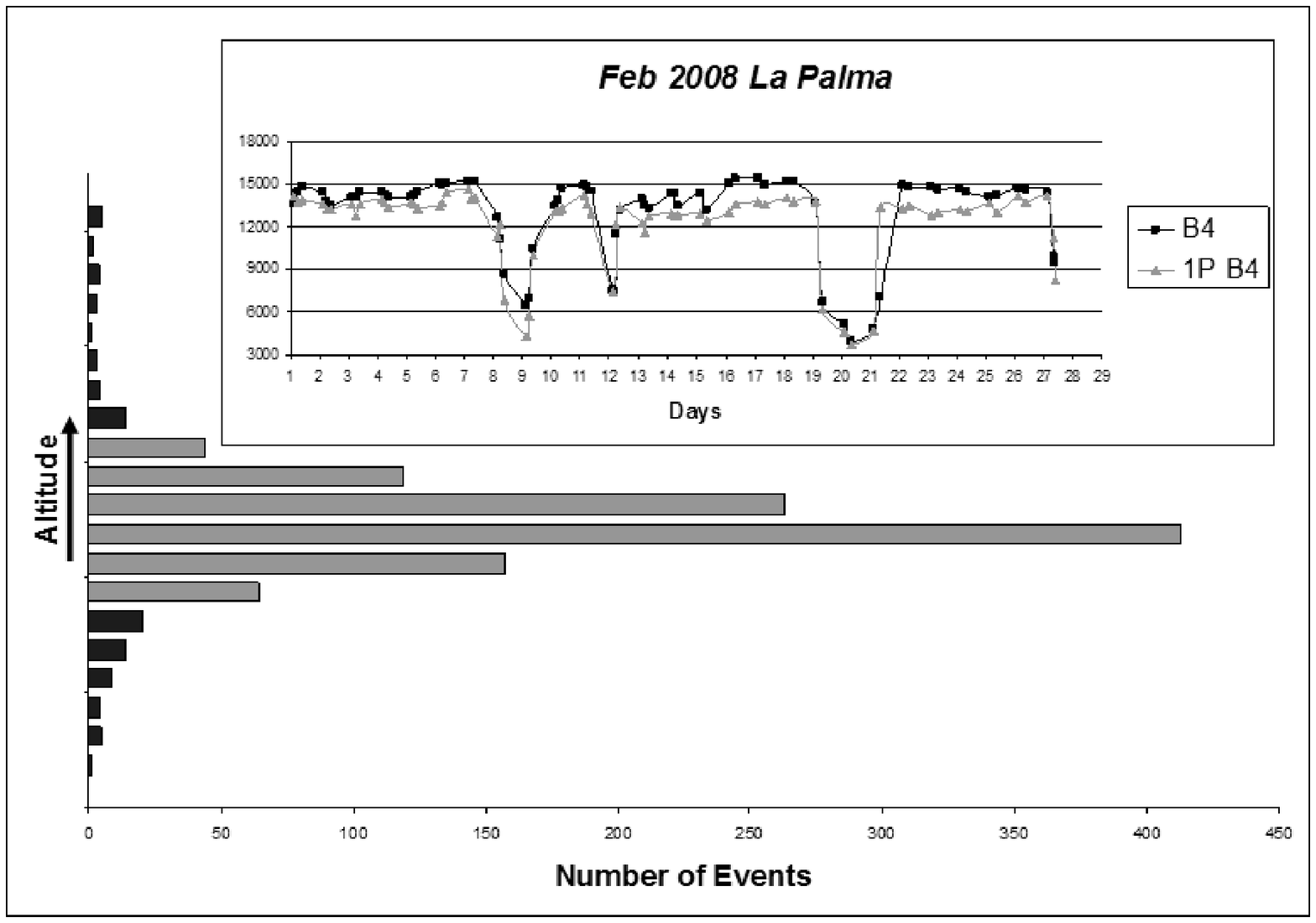}
   \caption{Correlation between the Matrix $1^\circ\times1^\circ$ vs the single pixel. At La Palma in $2008$ the histogram shows a correlation of $92\%$. The gray bars represent the data within the threshold $\leq\left|1\sigma\right|$, while black bars are the data with difference $>\left|1\sigma\right|$. The upper panel represents a pattern of $1^\circ\times1^\circ$ matrix (black line) and the single pixel (gray line) in B4 band for a single month. The altitude corresponding to the peak of the histogram corresponds to about $4000m$ as indicated by the B4 weighting function of GOES12 satellite \textbf{(Bin$=250m$)}. We note that the main differences arise from low-altitude phenomena.}
             \label{1pixel}
   \end{figure}

\begin{figure}
  \centering
  \includegraphics[width=8.5cm]{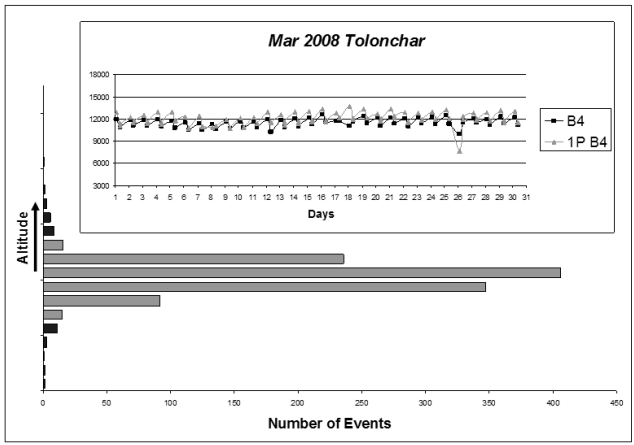}
  \caption{Correlation between the Matrix $1^\circ\times1^\circ$ vs the single pixel. At Tolonchar in $2008$ the histogram shows a correlation of $96\%$. The gray bars represent the data within the threshold $\leq\left|1\sigma\right|$, while black bars the data with difference $>\left|1\sigma\right|$ \textbf{(Bin$=250m$)}. The upper panel represents a pattern of $1^\circ\times1^\circ$ matrix (black line) and the single pixel (gray line) in B4 band for a single month.}
             \label{1pixelt}
   \end{figure}

\section{Satellite data acquisition}
\label{satacqui}

For the purposes of this work, we used GOES12 equipped with the imager. Among the 5 available channels, as shown in Table \ref{BAND}, we have selected the  water vapor channel (channel 3, hereafter called B3 band) centered at $6.7~\mu m$, the cloud coverage channel (channel 4, hereafter called B4 band) centered at $10.7~\mu m$, and the $CO_2$ band (channel 6, hereafter called B6 band) centered at $13.3~\mu m$. B3 band is sensitive between $6.5-7.0~\mu m$ and is able to detect high altitude cirrus clouds, B4 band is sensitive between $10.2-11.2~\mu m$ and is able to detect middle level clouds, while B6 band is able to sense  small particle such as fog, ash and semi-transparent high clouds.
Data are  a measurements of thermal radiation emitted during the night.
The selection of the IR channels was done in order to detect clouds at different heights, because water vapor absorbs electromagnetic radiation and then re-emits it in various wavelength bands, in particular in the infrared region at $6-7~\mu m$. If clouds are not present, the emissions at  $10.7~\mu m$ reaching the satellite is largely not absorbed  by the atmosphere so the measured radiance values are due to emission from surface. Instead when clouds are present, the emissivity drops.  Data are prepared by the Comprehensive Large Array-data Stewardship System (CLASS), an electronic library of NOAA environmental data\footnote{www.class.ngdc.noaa.gov}, and are stored as rectified full earth disk images in a format called AREA files.
We processed them using McIDAS-V Version 1.0beta4, a free ware software package.
First we extracted the GOES data on the telescope sites.\\Table \ref{ST} shows the geographic coordinates of the analysed sites. For each site we have identified and extracted  a sub-image of $1^\circ \times 1^\circ$ having the central
pixel centered on (or near) the  coordinates given in Table \ref{ST}.\\Due to the discrete grid of the available GOES data the distances from the central pixel for each site are: $5'\pm1'$ at Paranal, $6'\pm1'$ at La Silla, $5'\pm1'$ at La Palma, $3'\pm1'$ at Mt. Graham and $4'\pm1'$ at Tolonchar. These distances are very small compared to the used matrix.\\
For each night we have extracted the observations at three different hours: at 02:45, 05:45, 8:45 because they are the local times in common for all sites under investigation available from GOES12 satellite. In case of not availability of the specific images, the nearest temporal image was used. The last column of Table \ref{ST} shows the satellite view angle. Figure \ref{CMAT} shows the two different projections obtained from each acquisition at La Palma and Tolonchar.

\begin{table}
 \centering
 \begin{minipage}{80mm}
  \caption{Geographic characteristics of the analyzed sites and GOES12 satellite. The view angle is obtained through the formula $\theta=\sqrt{(\Delta LAT)^{2}+(\Delta LONG)^{2}}$.}
   \label{ST}
  \begin{tabular}{@{}lcccc@{}}
  \hline

  site      &LAT.&      LONG. & Altitude & View Angle \\
            &    &            &  Km      &            \\
 \hline
 Paranal    &   $-24^{\circ}37'$  &  $-70^{\circ}24'$  &  $2.630$   &  $25^{\circ}00'$  \\
 La Silla   &   $-29^{\circ}15'$  &  $-70^{\circ}43'$  &$2.347$     &  $29^{\circ}30'$    \\
 La Palma   &   $+28^{\circ}45'$  &  $-17^{\circ}52'$  &  $2.363$   &  $64^{\circ}10'$   \\
 Mt.Graham  &   $+32^{\circ}42'$  &  $-109^{\circ}52'$ &  $3.267$   &  $47^{\circ}40'$  \\
 Tolonchar  &   $-23^{\circ}56'$  &  $-67^{\circ}58'$  & $4.480$    &  $24^{\circ}50'$  \\
 \hline
 GOES12     &   $+0^{\circ}00' $  &  $-75^{\circ}00'$  & $35800$    &                   \\
 \hline

\end{tabular}
\end{minipage}
\end{table}

In Paper III the analysis of the amount of clear sky fraction at La Palma and Mt.Graham was based following the same approach of Erasmus \& van Rooyen (\cite{erasmus06}). We have used the B3 and B4 bands separately to sense thick clouds, but the old procedure presented some limits in case of partial coverage or thin clouds. In this paper we refine the analysis using a new and more sophisticated channel correlation analysis in order to detect more subtle effects due to atmospheric perturbations, including sudden changes in air masses, which imply changes in seeing, wind and relative humidity. We also included in the analysis the B6 band, see Section \ref{analysis}. We believe that these previous limits are overcame by correlating B4 with B3 and B6 bands.
Another difference in this new analysis is that the flux is averaged on an area of $1^\circ \times 1^\circ$ instead the 1 pixel value obtaining significant decrease of the instrumental noise. A comparison of the two procedures is described in the following section.
This matrix analysis is validated using the GOES 12 data of La Palma and Tolonchar because the two sites show very different geophysical conditions and different satellite angle of view. After the positive validation we decided to extend the same analysis to Paranal, La Silla, and Mt.Graham.

\subsection{Resolution Correlation Matrix}
\label{rcm}

The histograms of Figures \ref{1pixel} and \ref{1pixelt} show the correlation between the Matrix $1^\circ\times1^\circ$ vs the single pixel at La Palma and Tolonchar in the year  2008 for B4 band. Section \ref{epatd} describes the selected threshold used in this classification.\\ The grey bar of each histogram represents the data with absolute value $\leq\left|1\sigma\right|$ level, black bars show data $>\left|1\sigma\right|$ level. Moreover the histogram  represents the distribution of the B4 band in altitude. The peak of the histogram corresponds to about $4000m$ at La Palma, and greater than $4000m$ at Tolonchar, see Section \ref{WFS}.\\ The altitude was extrapolated from the B4 weighting function of GOES12 satellite. The peaks of these functions at high altitude and the use of matrix make the model suitable and sensitive for the study of atmospheric layers above the telescope sites. In particular, clouds below the level of the observing site do not affect the model as demonstrated by the high correlation percentage.\\
In fact the $92\%$ of data at La Palma are within $\left|1\sigma\right|$ level, the $96\%$ at Tolonchar.
The inset plots of the Figures \ref{1pixel} and \ref{1pixelt} represent the correlation of B4 flux computed as a mean value in a $1^\circ\times1^\circ$ matrix (black line), and B4 flux obtained in a single pixel (grey line) for March 2008 at Tolonchar and February 2008 at La Palma. February is chosen as a typical perturbed month because of the wide count fluctuations. In each case we see that the mean matrix and 1 pixel values show a similar pattern, this means that we are looking at a high altitude compared to the height site. For the figures we have chosen critical months to show that the correlation is good for any atmospheric condition and then of the season.\\
We decide to use the matrix, instead the 1 pixel value, because the average of the flux gives more stable information reducing the fluctuations due to instrumental noise.  A further advantage in the use of matrix is that we are looking at a wider field of view than in the one-pixel analysis.\\
We are confident that thanks to the high correlation we obtain statistical reliable data. As shown, the histogram is correlated with the altitude of the site, through the comparison between the matrix and the single pixel we can also extract information on the site analyzed.\\ La Palma histogram clearly shows an asymmetric distribution showing that perturbations are mainly due to low altitude.
A check we done to confirm this point extracting  the 
log comments of data located on the low side of the histogram queue. We 
found that the majority of the comments are "freezing fog" and data are 
from winter time.\\ At Tolonchar the distribution is symmetric and with an almost negligible queue.

\section{Remote Sounding Basic Model}
\label{rsbm}

The mathematical model used in this analysis is here explained. The emitted monochromatic radiation intensity at a given $\lambda$ and
along a vertical path at the top of the atmosphere, incident at a satellite instrument is given by:

\begin{equation}
	R_{\lambda}=(I_{0})_{\lambda}\tau_{\lambda}(z_{0})+\int^{\infty}_{z_{0}}B_{\lambda}{T(z)}K_{\lambda}(z)dz
	\label{eq:a}
\end{equation}

where:

\begin{itemize}
	\item $K_{\lambda}(z)=\frac{d\tau_{\lambda}(z)}{dz}\Rightarrow$ Weighting Function (WF)
	\item $B_{\lambda}{T(z)}\Rightarrow$ Planck function profile as function of vertical temperature profile T
	\item $(I_{0})_{\lambda}\Rightarrow$ Emission from the earth surface at height $z_{0}$
	\item $\tau_{\lambda}(z)\Rightarrow$ Vertical transmittance from height $z$ to space
\end{itemize}

This equation may also be extended to represent radiation emitted along a slant (non-vertical) path making the approximation of a plane-parallel atmosphere.\\ For a viewing path through the atmosphere at angle $\theta$ to the vertical, we have:

\begin{equation}
	\tau_{\lambda}(z,\theta)=e^{-sec\theta\int^{\infty}_{z}K_{\lambda}(z)c(z)\rho(z)dz}	
\end{equation}

where:

\begin{itemize}
	\item $\rho(z)\Rightarrow$ Vertical Profiles of Atmospheric Density
	\item $K_{\lambda}(z)\Rightarrow$ Absorption Coefficient
	\item $c(z)\Rightarrow$ Absorbing Gas Mixing Ratio
\end{itemize}

Changing from the notation of a continuous profiles, as in equation (\ref{eq:a}), to discrete profile, and considering the atmosphere as a composition of many thin layers, the corresponding equation becomes:

\begin{equation}
	R_{i}=(I_{0})_{i}\tau_{i}(z_{0})+\sum^{j-1}_{j=1}B_{ij}K_{ij}
\end{equation}

Making substitutions as below:

\begin{enumerate}
	\item $B_{j}\Rightarrow I_{0}$
	\item $K_{ij}\Rightarrow \tau_{i}(z_{0})$
\end{enumerate}

Hence equation is given by

\begin{equation}
	R_{i}=\sum^{j}_{j=1}B_{j}K_{ij}	
\end{equation}

Representing the radiance in all channels and the Planck function profile by vectors we have:

\begin{equation}
	\vec{R}=\vec{B}\cdot\vec{K}
\end{equation}

where $\vec{K}$ is a matrix containing the discrete weighting function elements $i\times j$. Assuming the problem to be linear (i.e. $\vec{K}$ is $\vec{B}$ independent) the  formula to find out the $\vec{B}$ function can be inverted.

\subsection{The Weighting Functions of GOES12}
\label{WFS}

The weighting function (WF) specifies the layer from which the radiation emitted to space originates, and hence it determines the region of the atmosphere which can be sensed from space at fixed $\lambda$.\\ In such a way many atmospheric layers can be observed by selecting different $\lambda$ values.\\ If a standard atmosphere is assumed GOES12 WFs have the following median height values\footnote{See http://cimss.ssec.wisc.edu/}:

\begin{itemize}
	\item BAND3: $K_{\lambda_{3}}(z)=\frac{d\tau_{\lambda_{3}}(z)}{dz}\Rightarrow \approx 8000m$
	\item BAND4: $K_{\lambda_{4}}(z)=\frac{d\tau_{\lambda_{4}}(z)}{dz}\Rightarrow \approx 4000m$
	\item BAND6: $K_{\lambda_{6}}(z)=\frac{d\tau_{\lambda_{6}}(z)}{dz}\Rightarrow \approx 3000m$
\end{itemize}

These heights depend on the location of the selected earth region. For instance Tolonchar B3 height is supposed to be fairly constant while B4 and B6 heights are higher because the site is $4480m$ height. In any case experimental observations confirm that GOES12 B3-B4-B6 bands looks at high layers as regard to soil.

\section{The analysis of the infrared B3-B4-B6 GOES12 Bands}
\label{analysis}

A cloud cover analysis is possible by mean of Remote Sounding (RS) application to B3, B4 and B6: in the current models based mainly on B4 analysis only. This band fairly matches thick cloud observation, but it presents some limits in case of thin clouds or minor atmospheric events. These limits are mostly overcome by correlating this band with B3 and B6. Ground vs satellite data show that B3 is capable of detecting atmospheric events such as winds or relevant air displacements. Moreover different air mass changes (e.g. dry, wet, warm or cold wind) is detectable by comparing B4 to B3. Finally a correlation between B6 and B4 allows to gather information about fogs, dusts, thin clouds. In such a way remote sounding model applied to GOES12 bands provides the following atmospheric scheme:

\begin{itemize}
	\item B3-B4 correlation: high atmospheric events and in particular air mass displacements.
	\item B4-B6 correlation: low atmospheric events, and in particular fogs, dusts, humidity.
\end{itemize}

\begin{figure}
  \centering
  \includegraphics[width=8.5cm]{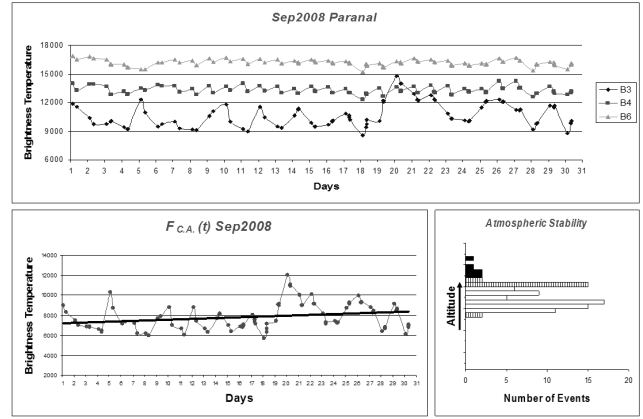}
  \caption{GOES 12 emissivity in B3, B4, B6 bands (upper panel) at Paranal for September 2008. Left panel shows
  the correlation function (the black straight line represents the $F_{C.A.}(t)$ trendline).
   The corresponding atmospheric stability histogram is shown in the right lower panel.}
             \label{a2}
\end{figure}

On this base we can provide a sort of  \textbf{atmospheric tomography} by satellite data extrapolation. Figure \ref{a2}  shows the distribution of GOES12 emissivity in the three bands  at Paranal (upper panel) and the distribution of correlation function $F_{C.A.}(t)$ (bottom left panel) for the month of September 2008. The corresponding atmospheric correlation function it is also shown in the right side of the panel. For each month we have obtained these distributions.

\begin{figure}
  \centering
  \includegraphics[width=8.5cm]{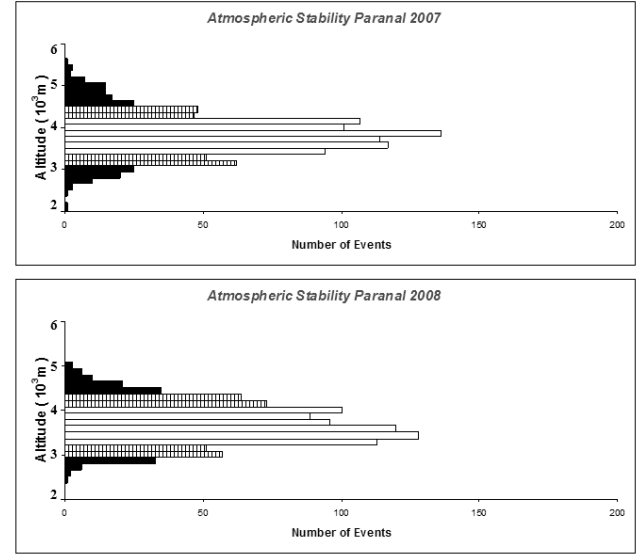}
  \caption{Histogram of annual atmospheric stability at Paranal. White bars represent
the stable nights, gray bars clear but unstable nights, black
bars the nights covered.}
             \label{s}
   \end{figure}

\begin{figure}
  \centering
  \includegraphics[width=8.5cm]{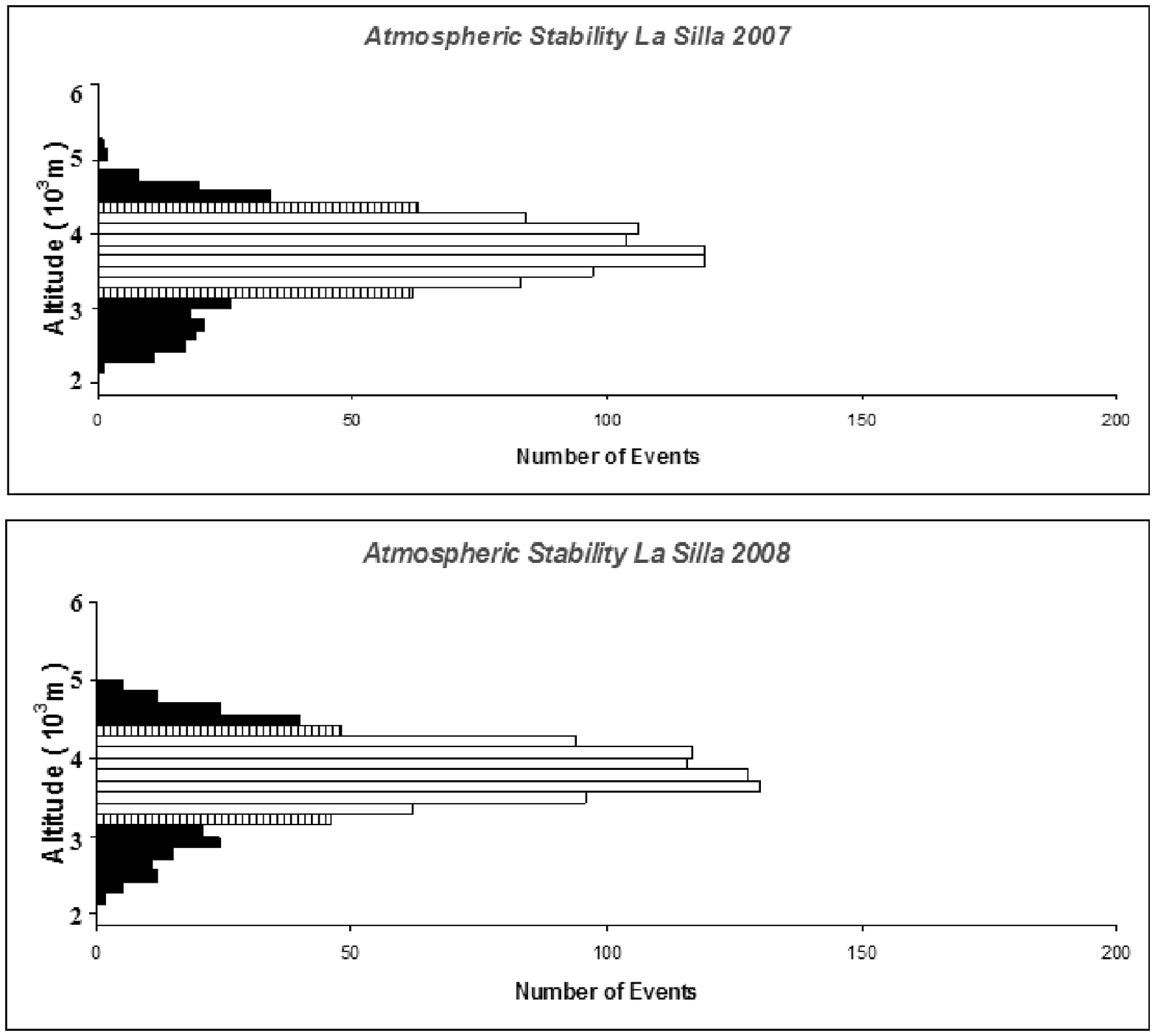}
  \caption{Histogram of annual atmospheric stability at La Silla. White bars represent
the stable nights, gray bars clear nights but unstable, black
bars the nights covered.}
             \label{s1}
   \end{figure}

\begin{figure}
  \centering
  \includegraphics[width=8.5cm]{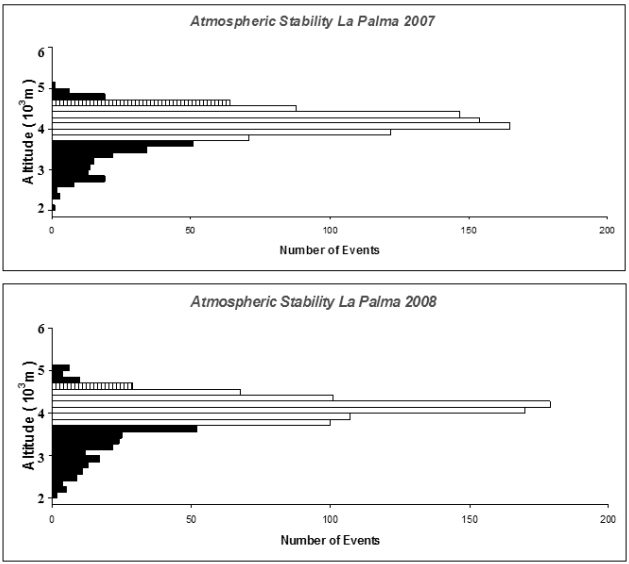}
  \caption{ Histogram of annual atmospheric stability at La Palma. White bars represent
the stable nights, gray bars clear nights but unstable, black
bars the nights covered. We note that La Palma instability (black bars) is due mainly to low-altitude phenomena such as
fog, dust, etc. as confirmed by log comments.}
             \label{s2}
   \end{figure}

\begin{figure}
  \centering
  \includegraphics[width=8.5cm]{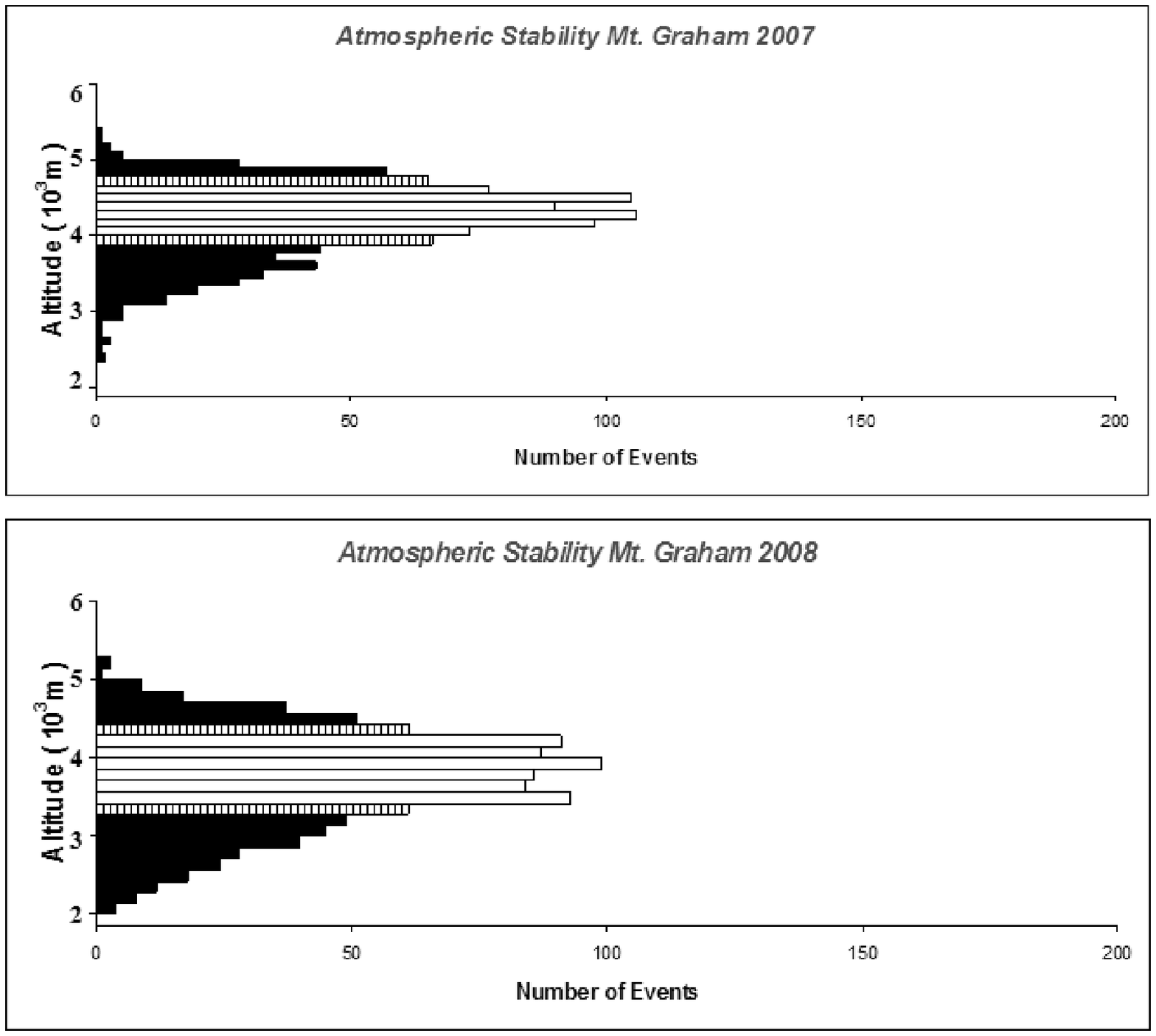}
  \caption{Histogram of annual atmospheric stability at Mt.Graham. White bars represent
the stable nights, gray bars clear nights but unstable, black
bars the nights covered.}
             \label{s3}
   \end{figure}

\begin{figure}
  \centering
  \includegraphics[width=8.5cm]{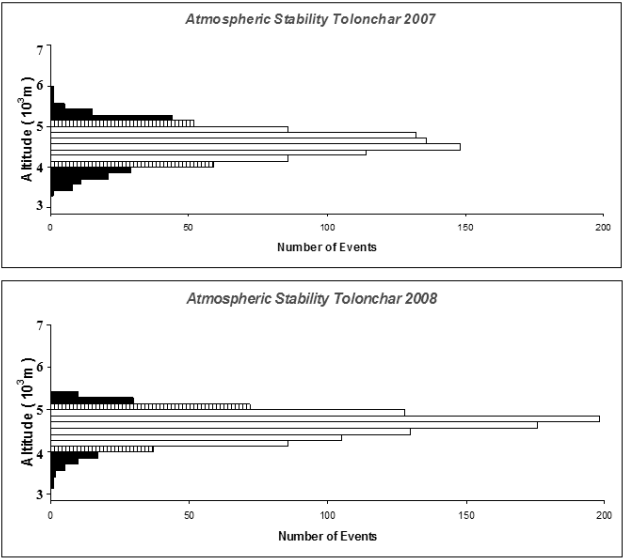}
  \caption{Histogram of annual atmospheric stability at Tolonchar. White bars represent
the stable nights, gray bars clear nights but unstable, black
bars the nights covered.}
             \label{s4}
   \end{figure}

\section{Atmospheric Correlation Function}
\label{acf}

As written in the previous sections the $F_{C.A.}(t)$ correlation function used in this analysis is based on three band correlation. Considering GOES 12 WFs the best Remote Sounding is:

\begin{equation}
I_{C.A.}=I_{\lambda_{3}}-[I_{\lambda_{6}}-I_{\lambda_{4}}]
\end{equation}

This model takes into account auto-corrections of atmosphere: for instance if two high layers have a positive oscillation and lower layers has an equal magnitude oscillation, but negative, the $F_{C.A.}(t)$ remains constant. From the physical point of view this means that the corresponding wave front observed from earth will be automatically corrected and an event is never been observed, as high atmosphere oscillations provoked by the B3-B4 correlation are always greater than those provoked by the B4-B6 correlation. B4-B6 oscillations can only partially correct the wave front.\\
In mathematical terms this model provides a brightness temperature of the B3, B4 and B6 combination, given by equation:

	\[I_{C.A.}=\frac{R_{\lambda_{3}}+R_{\lambda_{4}}-R_{\lambda_{6}} }{\tau(z_{0})}+
\]

\begin{equation}
-\frac{\int^{\infty}_{z_{0}}B_{\lambda_{3}}[T(z)]K_{\lambda_{3}}+B_{\lambda_{4}}[T(z)]K_{\lambda_{4}}-B_{\lambda_{6}}[T(z)]K_{\lambda_{6}}dz}{\tau(z_{0})}	 
\end{equation}

$F_{C.A.}(t)$ can be extrapolated by relating T brightness to time. This function will provide information about atmospheric quality of the surveyed site and the height of the perturbation that is a
function of the T brightness.
We show in Section \ref{correspondence} that this function is related to the seeing.\\
Subsequently the air mass displacements (dynamical atmospheric instability) can be ranked
according to their altitude and then to the kinetic energy.
Figure \ref{a2} will represent a pattern of B3, B4 and B6 at Paranal.
The bottom of Figure \ref{a2} shows the correlation function
extrapolated through the RS and the respective histogram of
atmospheric stability. This histogram also gives us information on
the share that generates the disturbance. We note that the flat
distribution of the B4 and B6 show no cloud cover, unlike that of B3, which has
strong oscillations. Assuming a standard atmosphere for this band
observed at a height of $\approx8000m$, we infer that the phenomena
are of high altitude. Observations from the ground confirm the
presence of strong winds and a worsening of the seeing. Figures
\ref{s}, \ref{s1}, \ref{s2}, \ref{s3}, \ref{s4} depict the
histogram of annual atmospheric stability. White bars represent
the stable nights, gray bars clear nights but unstable, black
bars the nights covered (see Section \ref{csnc} for definitions). The thresholds were obtained solely from
analysis of satellite data. Histograms were derived from the
correlation function, so they can give information on the
contribution of atmospheric phenomenon in question.

\subsection{Satellite Atmospheric Tomography}
\label{sat}

The atmospheric stability is derived from the atmospheric correlation function.
This function, extrapolated from the RS of B3, B4 and B6 bands, is
correlated to the integrated structural parameter of the refraction index ($C^{2}_{n}$) and than to
the optical turbulence. In fact, as shown belove, the RS of the B3,
B4 and B6 bands is function of the changes in temperature at various
altitudes. Atmospheric stability can be estimated by involving the model depicted below.
Let's consider brightness temperature processed using \textbf{McIDAS-V} package. From a theoretical point of view the brighter a GOES12 image pixel  the hotter the observed layer (i.e. lower layer).\\
Emitted radiation intensities at different satellite observation ${\lambda}$ are then:

\[	 R_{\lambda_{3}}=(I_{0})_{\lambda_{3}}\tau_{\lambda_{3}}(z_{0})+
\int^{\infty}_{z_{0}}B_{\lambda_{3}}{T(z)}K_{\lambda_{3}}(z)dz
\]
	
	\[	 R_{\lambda_{4}}=(I_{0})_{\lambda_{4}}\tau_{\lambda_{4}}(z_{0})+
\int^{\infty}_{z_{0}}B_{\lambda_{4}}{T(z)}K_{\lambda_{4}}(z)dz
\]

		\[	 R_{\lambda_{6}}=(I_{0})_{\lambda_{6}}\tau_{\lambda_{6}}(z_{0})+
\int^{\infty}_{z_{0}}B_{\lambda_{6}}{T(z)}K_{\lambda_{6}}(z)dz
\]

These equations provide informations about layer height and temperature. Plotting these data as function of time an atmospheric instability function can be extrapolated.

\subsection{Correspondence between the Seeing and the Atmospheric Correlation Function}
\label{correspondence}

\begin{figure}
  \centering
  \includegraphics[width=8.5cm]{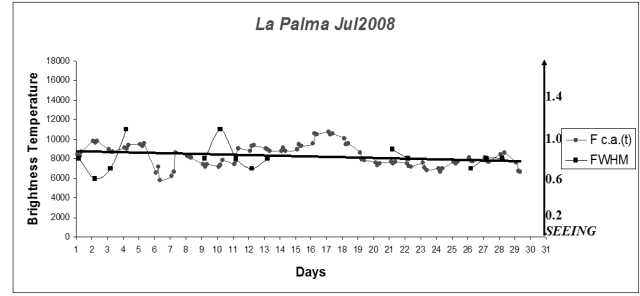}
  \caption{Atmospheric Correlation Function-FWHM Correspondence. La Palma, July 2008. The solid gray line represents $F_{C.A.}(t)$ trend. The discontinuous black line represents the available seeing values. The black straight line represents the $F_{C.A.}(t)$ trendline. We note that the worse seeing occurs when the \textit{MAX} and \textit{min} values of the $F_{C.A.}(t)$ correspond (Correlation Coefficient$=0.92$).}
             \label{FW7}
   \end{figure}

\begin{figure}
  \centering
  \includegraphics[width=8.5cm]{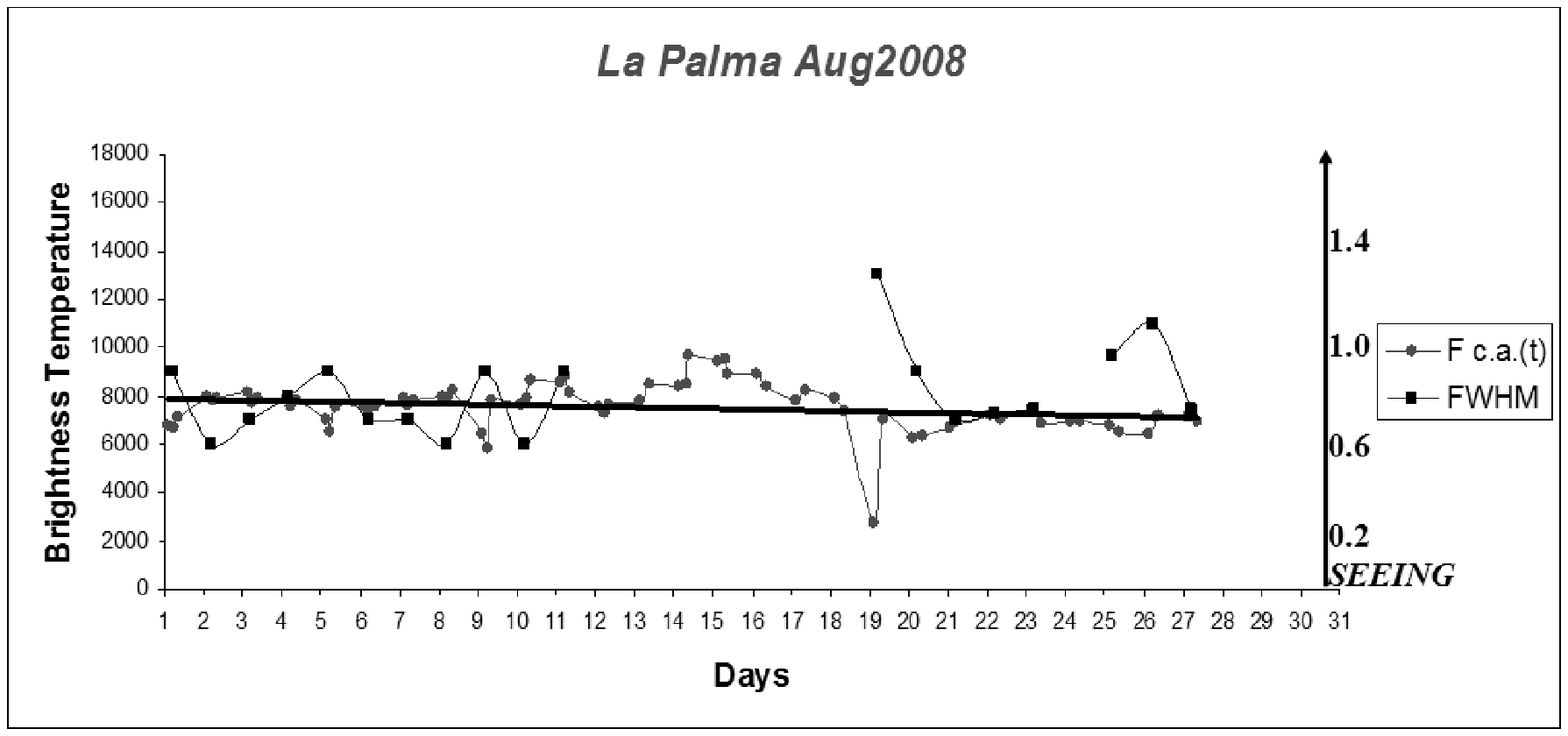}
  \caption{Atmosferic Correlation Function-FWHM Correspondence. La Palma, August 2008. The solid gray line represents $F_{C.A.}(t)$ trend. The discontinuous black line represents the available seeing values. The black straight line represents the $F_{C.A.}(t)$ trendline. We note that the worse seeing occurs when the \textit{MAX} and \textit{min} values of the $F_{C.A.}(t)$ correspond (Correlation Coefficient$=0.91$).}
             \label{FW8}
   \end{figure}

\begin{figure}
  \centering
  \includegraphics[width=8.5cm]{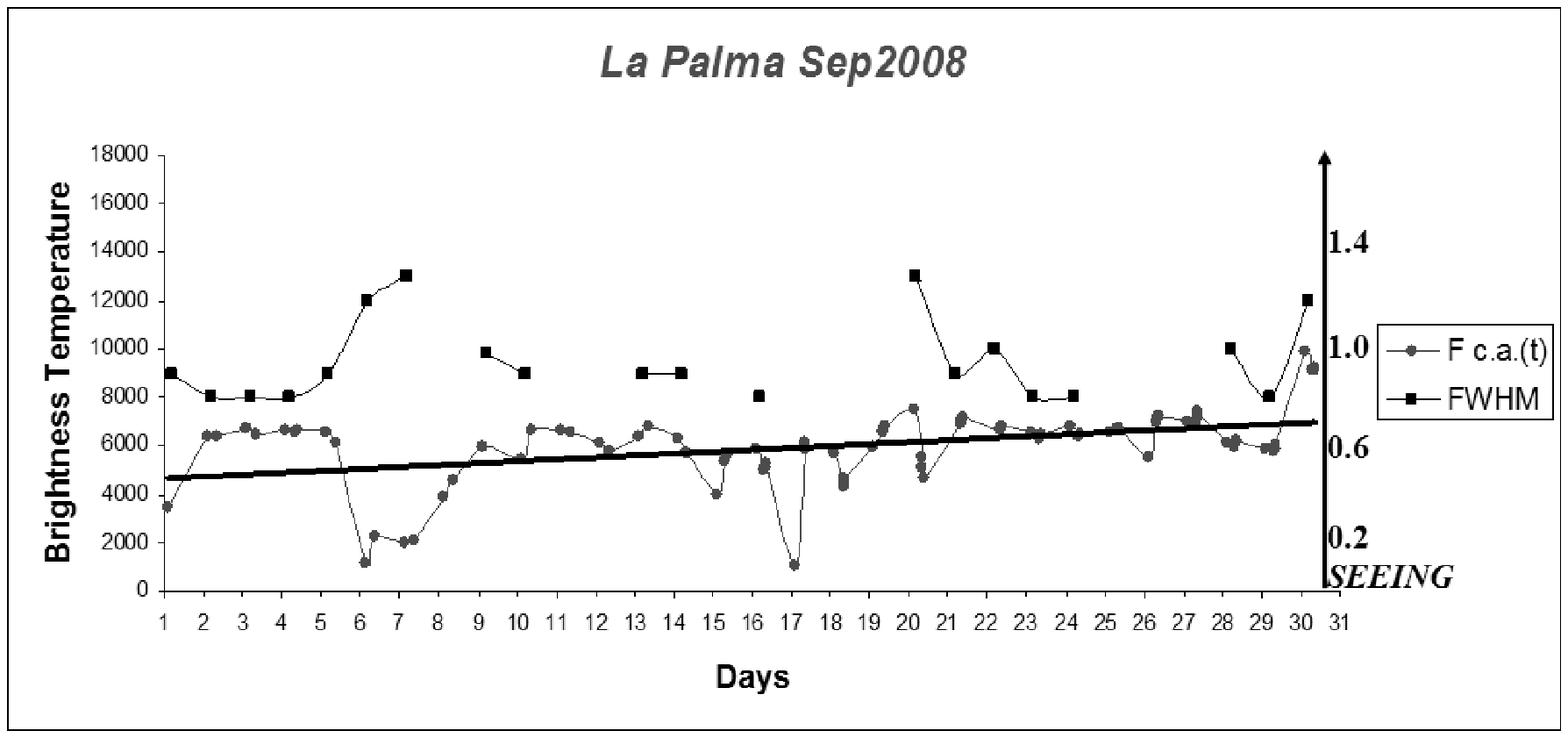}
  \caption{Atmosferic Correlation Function-FWHM Correspondence. La Palma, September 2008. The solid gray line represents $F_{C.A.}(t)$ trend. The discontinuous black line represents the available seeing values. The black straight line represents the $F_{C.A.}(t)$ trendline. We note that the worse seeing occurs when the \textit{MAX} and \textit{min} values of the $F_{C.A.}(t)$ correspond (Correlation Coefficient$=0.88$).}
             \label{FW9}
   \end{figure}

In this section we detect a possible correlation between the seeing
obtained from the web page of the Robotic Differential Image Motion
Monitor (known as RoboDIMM\footnote{See http://catserver.ing.iac.es/robodimm/}) of Isaac Newton Telescope (INT) and the atmospheric
correlation function $F_{C.A.}(t)$ computed for the la Palma sky to test a possible correlation with the image quality.\\
This RoboDIMM, like all classical DIMMs, relies on the method of
differential image motion of telescope sub-apertures to calculate the
seeing Fried parameter $r_{0}$. RoboDIMM forms four separated images of the same
star, and measures image motion in two orthogonal directions from which it
derives four simultaneous and independent estimates of the seeing. The data
intepretation makes use of the Sarazin and Roddier's DIMM
algorithm as described in (Sarazin \& Roddier (\cite{dimm})), based on
the Kolmogorov theory of atmospheric turbulence in the free atmosphere.
There is the possibility that some DIMMs, including the RoboDIMM may have
a lower limit threshold in the measurement of the seeing, due to noise,
but in our sample (Figures \ref{FW7}, \ref{FW8} and \ref{FW9}) the seeing values do not have values
significantly below $1 arcsec$. Moreover in this paper  we do not intend to
give an absolute calibration but only a correlation analysis of these two
functions.\\
In fact the solid gray line represents the $F_{C.A.}(t)$ trend. The discontinuous black line represents the available seeing values. We note that the seeing is worse if the $F_{C.A.}(t)>\left|1\sigma\right|$ as referred in Section \ref{csnc}.
A dedicated \textbf{Site Testing} can clearly improve current models, providing information about fundamental parameters. In a future paper we are planning, after an accurate set up of the ING's RoboDIMM (Isaac Newton Group of Telescopes), to correlate the values of seeing with the  values of $F_{C.A.}(t)$.\\
We specify that this is still a preliminary work. We plan in future to improve this model with the use of other seeing data bases and/or preferably $C^{2}_{n}(h)$ profiles.

\begin{figure}
  \centering
  \includegraphics[width=8.5cm]{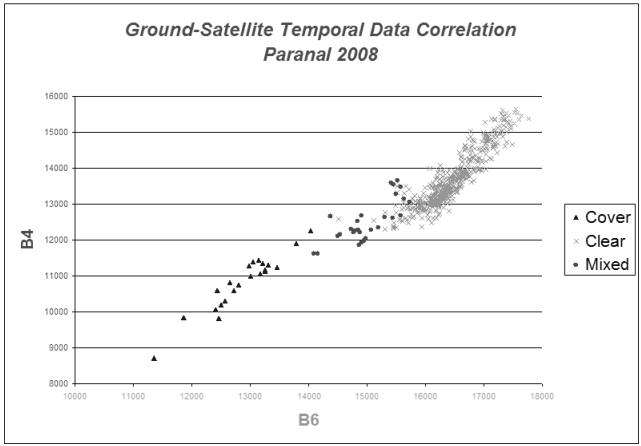}
  \caption{Temporal distribution of GOES12 B4 and B6 band emissivity  at Paranal in  2008.
  Sky quality classification has been carried out using the Paranal log.}
             \label{cor1}
   \end{figure}

\begin{figure}
  \centering
  \includegraphics[width=8.5cm]{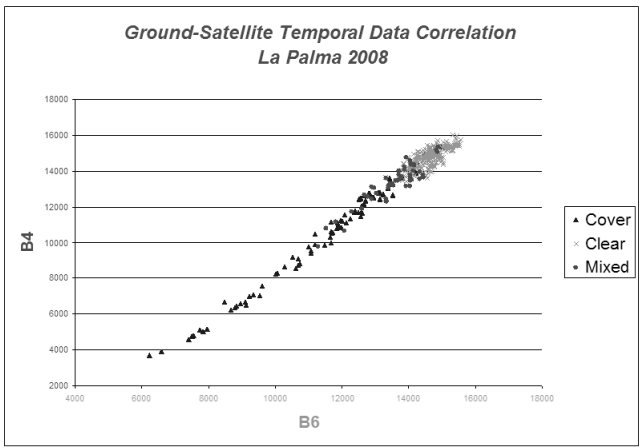}
  \caption{Temporal distribution of GOES12 B4 and B6 band emissivity at La Palma in  2008.
  Sky quality classification has been carried out using the merge of TNG and Liverpool ground based data.}
             \label{cor2}
   \end{figure}

\begin{figure}
  \centering
  \includegraphics[width=8.5cm]{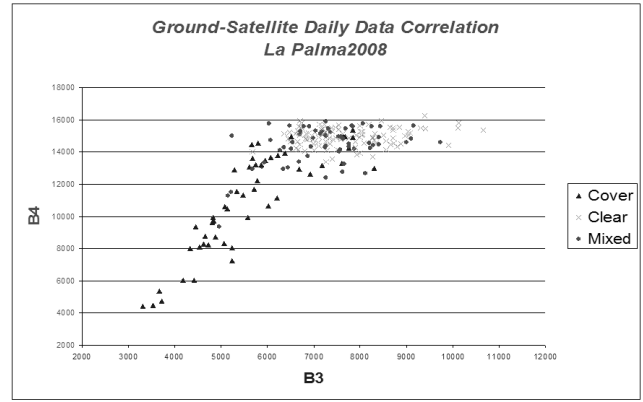}
  \caption{Daily distribution of GOES12 B3 and B4 band emissivity at La Palma in  2008.
  Sky quality classification has been carried out using the TNG log.}
             \label{cor3}
   \end{figure}

\section{Temporal Data Analysis}
\label{hda}

The ground-satellite correlation model used in this article is based on a temporal data correspondence of $\approx3$ values for each night.
We have in fact multiple values for each night ($\approx3$) and this gives us the opportunity to do a detailed analysis
of various conditions. We have also a larger number of data and this, from a statistical point of view, allows to
validate the model.\\
Figure \ref{cor1} and \ref{cor2} shown the plots of the obtained temporal emissivity of B4 band vs B6 for the 2008 at Paranal and at La Palma. The nights are classified  as a function of the sky quality obtained from the observing log of each analysed site. It appears that clear nights present high values of emissivity.\\ Moreover Fig. \ref{cor2} shows a lower dispersion and a better separation of clear nights if compared with Fig. \ref{cor3} confirming once more the better quality of the adopted method.
From Fig. \ref{cor1} and Fig. \ref{cor2} we can define as clear  all the nights having B4 $\geq13000~Units$  at Paranal, while at La Palma we can define as clear all the nights having  B4 $\geq13900~Units$.
Table \ref{delta} shows the obtained percentage of clear, mixed and covered nights at Paranal and at La Palma in the year 2008 from a temporal punctual analysis. As in Paper III, we have found that the fractions of clear time based on satellite data are greater than those of clear nights using ground based data.
The differences seem higher than the biases of single
logbooks. In fact the amount of nights computed  using the different
adopted logbooks gives a similar percentage with differences around 2\%
(computed in the same period, 2008-2009) when a careful and homogeneous
analysis is performed.\\ 
The obtained percentages are in reasonable agreement with the results reported in Garcia-Gil et al. (\cite{garcia}).\\
We can conclude that the nights classification is
more dependent on the adopted methodology and accuracy rather than on
biases in the adopted logbooks. The differences can be better explained considering that some local effects could be ignored by satellites.
In addition, we note that considering the fraction of clear time fraction from satellite data  vs. that of clear nights from ground data, the first fraction is obviously higher.\\ However we don't expect a large discrepancy as we demonstrated in Paper III where we found that the fraction of partially used nights is very small. It is interesting to note that we found an amount of satellite clear nights close to that obtained in Paper III even if we have used different bands: we obtained $71.9\%$ of clear nights in Paper III, obtained analysing  B3 vs B4 bands,  and $71\%$ of clear nights in the present analysis obtained plotting B4 vs B6 bands, but the thresholds are fixed considering only B4 (see Section \ref{cmcnc}).\\
The last row of Table \ref{delta} shows the percentage of accuracy to associate to each obtained fraction of nights. The uncertainty is computed as follows:

\begin{itemize}
	\item $\Delta_{Clear/Mixed}\Rightarrow$ Clear/Mixed Uncertainty
	\item $\Delta_{Clear/Covered}\Rightarrow$ Clear/Covered Uncertainty
	\item $\Delta_{Mixed/Covered}\Rightarrow$ Mixed/Covered Uncertainty
\end{itemize}

\begin{figure}
  \centering
  \includegraphics[width=8.5cm]{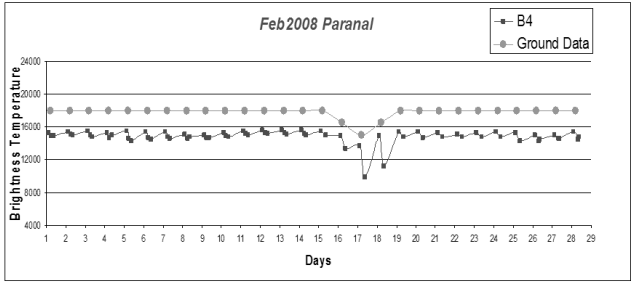}
  \caption{Ground Data-Satellite Data Correlation. Paranal, February 2008.}
             \label{corr}
   \end{figure}

We note that the largest satellite uncertainty derives from the overlap of clear and mixed nights, while the satellite
is accurate in other cases. It is interesting the comparison between the temporal and daily methods.\\
Figure \ref{cor3} represents daily distribution of GOES12 B3 and B4 band emissivity at La Palma in 2008. We note how the graph is more
dispersed compared to the graph of Figure \ref{cor2}. This is due to the greater accuracy of the temporal method and the
use of different bands: B3 and B6 respectively. In fact the B6 trend is more regular.\\
Table \ref{delta1} shows the comparison between temporal and daily data analysis. We can observe how the temporal method
uncertainties are smaller than the daily method uncertainties. In this case we chose annual thresholds and considered the
mathematical error, the method provides the greatest advantages choosing monthly thresholds and considering the statistical error.
The monthly thresholds make it possible to consider seasonal temperature changes of the site reducing the overlap percentages.\\
As final check Figure \ref{corr}  plots the B4 emissivity (black line) and log ground data (gray line) for February 2008 at Paranal. It is evident that the monthly distribution of the emissivity follows the ground data.

\begin{table*}
 \centering
 \begin{minipage}{170mm}
  \caption{Clear/Mixed/Covered nights percentage and overlaps at Paranal and La Palma in 2008. Temporal data analysis.}

   \label{delta}
  \begin{tabular}{@{}llllccccccc@{}}
  \hline
  &   \multicolumn{3}{c}{Ground} & \multicolumn{3}{c}{Satellite}\\
     & Clear& Mixed & Covered& Clear& Mixed& Covered\\

 \hline
 Paranal   & 91\% & 7\%    & 2\% & 84\% & 14\% & 2\% \\
 La Palma  & 66\%  & 12\%   & 22\% & 71\%   & 11\%   &   18\% \\
 \hline
  &  \multicolumn{3}{c}{Paranal} & \multicolumn{3}{c}{La Palma}\\
Uncertainty & $\Delta_{Clear/Mixed}$   &  $\Delta_{Clear/Covered}$   & $\Delta_{Mixed/Covered}$ & $\Delta_{Clear/Mixed}$   &  $\Delta_{Clear/Covered}$   & $\Delta_{Mixed/Covered}$  \\
\hline
Percentage &  $7\%$ &  $1\%$  &  $2\%$ &   $7\%$ &  $3\%$  &  $5\%$   \\
 \hline

\end{tabular}
\end{minipage}
\end{table*}

\begin{table*}
 \centering
 \begin{minipage}{170mm}
  \caption{Clear/Mixed/Covered nights percentage and overlaps at La Palma in 2008. Comparison between temporal and daily data analysis.}

   \label{delta1}
  \begin{tabular}{@{}llllccccccc@{}}
  \hline
  &   \multicolumn{3}{c}{Ground} & \multicolumn{3}{c}{Satellite}\\
     & Clear& Mixed & Covered& Clear& Mixed& Covered\\

 \hline
 La Palma (Daily)   & 60\% & 21\%    & 19\% & 69\% & 15\% & 16\% \\
 La Palma (Temporal) & 66\%  & 12\%   & 22\% & 71\%   & 11\%   &   18\% \\
 \hline
  &  \multicolumn{3}{c}{La Palma (Daily)} & \multicolumn{3}{c}{La Palma (Temporal)}\\
Uncertainty & $\Delta^{D}_{Clear/Mixed}$   &  $\Delta^{D}_{Clear/Cover}$   & $\Delta^{D}_{Mixed/Cover}$ & $\Delta^{T}_{Clear/Mixed}$   &  $\Delta^{T}_{Clear/Cover}$   & $\Delta^{T}_{Mixed/Cover}$  \\
\hline
Percentage &  $9\%$ &  $3\%$  &  $9\%$ &   $7\%$ &  $3\%$  &  $5\%$   \\
 \hline

\end{tabular}
\end{minipage}
\end{table*}

\subsection{ Discussion of Error Propagation and Thresholds}
\label{epatd}

\begin{figure}
  \centering
  \includegraphics[width=8.5cm]{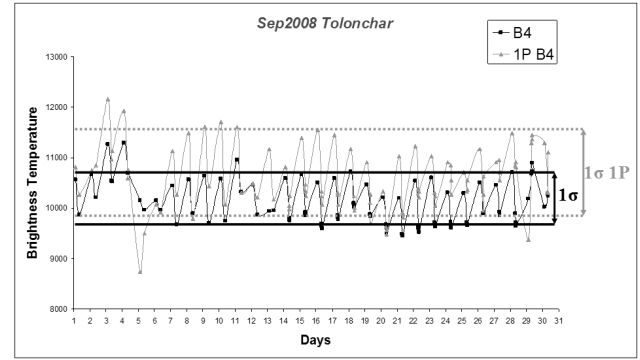}
  \caption{Comparison between the single pixel threshold of $1\sigma$ and the matrix threshold of $1\sigma$ at Tolonchar in September $2008$. Figure represents a pattern of $1^\circ\times1^\circ$ matrix (black line) and the single pixel (gray line) in B4 band for a single month. We note how the use of the matrix decreases the threshold value reducing the satellite noise. This makes the model more accurate.}
             \label{noise}
   \end{figure}

In our model the various thresholds to classify the nights were chosen by the individual analysis of satellite data. This also allows to study sites for which we have no ground data.\\
The thresholds were selected via the night temperature range detected by satellite and not through the real night brightness temperature range of the site.\\
This choice was made because the satellite temperature resolution decreases with the observation angle. In fact at La Palma we observe a temperature range lower than other sites. If we consider:

	\[Night~Satellite~Temperature~Range=1\sigma
\]

we note that the use of the matrix decreases the threshold value reducing the satellite noise. This makes the model more accurate (Fig.\ref{noise}).\\
The thresholds for each data classification are described below.

\subsubsection{Resolution Correlation Matrix Thresholds}
\label{rcmt}

In this section we compare the brightness temperature difference between the single pixel and the matrix method. We assume that the data are correlated if the difference is $\leq\left|1\sigma\right|$. Figures \ref{1pixel} and \ref{1pixelt} report the obtained percentage of correlation.

	\[\left|T^{1Pixel}_{Brightness}-T^{Matrix}_{Brightness}\right|<\left|1\sigma\right|
\]

where:

\begin{itemize}
	\item $T^{1Pixel}_{Brightness}\Rightarrow$ Brightness temperature of the single pixel
	\item $T^{Matrix}_{Brightness}\Rightarrow$ Brightness temperature of the $1^\circ\times1^\circ$ matrix
\end{itemize}

\subsubsection{Clear, Mixed, Covered Nights Classification}
\label{cmcnc}

The nights are classified using GOES12 B4 band. The classification of the nights is based in the following assumption: the maximum monthly brightness temperature $T^{Max}_{B}$ occurs in clear condition.\\
The other monthly brightness temperatures are correlated with  $T^{Max}_{B}$ when:

\begin{itemize}

	\item $T^{Max}_{B}-T_{B}\leq2\sigma\Longrightarrow$ Clear
	\item $2\sigma<T^{Max}_{B}-T_{B}\leq3\sigma\Longrightarrow$ Mixed
	\item $T^{Max}_{B}-T_{B}>3\sigma\Longrightarrow$ Covered
	
\end{itemize}

where $T_{B}\Rightarrow$ Brightness temperature of the $1^\circ\times1^\circ$ matrix.\\
From this definition "clear" sky means a matrix where there are no clouds. As concerning the ground based data we define "clear" the nights cloud free in the logbooks and completely usable for observations.\\
"Mixed" are nights where comments of presence of clouds or meteorological events (fog, wind, humidity...) have been found, but part of night was used.\\
"Covered" are unusable nights due to clouds or fog.

\subsubsection{Clear, Stable Nights Classification}
\label{csnc}

We calculated the monthly percentage of clear time relying on temporal data analysis. With this method we classify the fraction of each night by reading multiple data (e.g. if we have three data for a night, two clear values and one covered value, the percentage of clear night will be $67\%$). This is a definition close to the classical "spectroscopic time". We define "stable" a clear sky without atmospheric phenomena that may affect the photometric quality (wind, fog, humidity…).\\ 
The monthly percentage of photometric time is calculated by the same method and this is close to the classical definition of the "photometric time". This classification is very important because the photometric quality of clear sky, is influenced by phenomena not detectable by the methods currently used.\\
Finally, we clarify that an unstable sky might be still useful for observing because it is a subset of clear sky. This explains the differences in percentages of our classification (see Tables \ref{Mean07} and \ref{Mean08}).\\ 
To be more clear we specify that we calculated the time fraction, not the whole night fraction.\\If we take into account the atmospheric correlation function $T^{Max}_{B}$, it is possible to introduce the concept of stable nights.  Considering $F_{C.A.}(t)$ trendline we get the following classification:

\begin{itemize}
	\item $\left|T_{B}-T^{Trendline}_{B}\right|\leq\left|1\sigma\right|\Longrightarrow$ Stable
	\item $\left|1\sigma\right|<\left|T_{B}-T^{Trendline}_{B}\right|\leq\left|2\sigma\right|\Longrightarrow$ Clear
	\item $\left|T_{B}-T^{Trendline}_{B}\right|>\left|2\sigma\right|\Longrightarrow$ Covered
	
\end{itemize}

where:

\begin{enumerate}
  \item $T^{Trendline}_{B}\Rightarrow$ Brightness temperature of the monthly trendline
	\item $T_{B}\Rightarrow$ Brightness temperature of the $1^\circ\times1^\circ$ matrix
\end{enumerate}

Through this classification we obtain the histograms in Figures \ref{s}, \ref{s1}, \ref{s2}, \ref{s3}, \ref{s4}. White bars represent
the stable nights, gray bars represent clear but unstable nights, black
bars represent the covered nights.\\
The thresholds were obtained only from analysis of satellite data. Histograms were derived from the
correlation function, so they can give information on the
contribution of atmospheric phenomenon.
The final results are reported in Tables \ref{Mean07} and \ref{Mean08}.

\subsubsection{Mathematics Errors Propagation}
\label{mep}

Tables \ref{delta}, \ref{delta1} show the uncertainties to associate to each single data computed
through the formula:

	\[\Delta_{Tot}=\sqrt{(\Delta_{Cl/Mix})^{2}+(\Delta_{Cl/Co})^{2}+(\Delta_{Mix/Co})^{2}}\]

obtaining the following values:

\begin{itemize}
	\item Paranal $\Rightarrow \Delta_{Total}=7.3\%$
	\item La Palma (Temporal) $\Rightarrow \Delta_{Total}=9.1\%$
	\item La Palma (Daily) $\Rightarrow \Delta_{Total}=13.1\%$
\end{itemize}

We observe that the temporal method for La Palma reduces the total uncertainty by $4\%$.

\begin{figure}
  \centering
  \includegraphics[width=8.5cm]{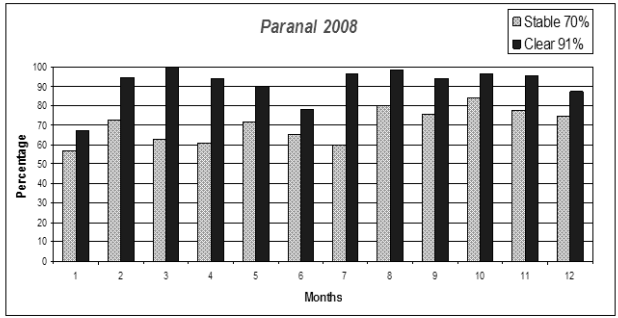}
  \caption{Clear and stable night fractions at Paranal 2008 from GOES12 satellite.}
             \label{p}
\end{figure}

\subsubsection{Statistics Errors Propagation}
\label{sep}

Now we consider the statistical error with the unbiased data assumption. We have $N(Ground;Satellite)$ pairs of values and in this case, considering the mathematical error for each site, the standard deviation on the function $F_{\Delta}(Ground;Satellite)$ is derived by the formula:

	\[\sigma_{F_{\Delta}}=\left[\Delta_{Total}\right]^{2}
\]

Finally, if we consider the number of data ($N(Ground;Satellite)$), the annual statistical uncertainty ($\Delta_{Statistical}$) of the model is given by the formula:

	\[\Delta_{Statistical}=\frac{\sigma_{F_{\Delta}}}{\sqrt{N(G;S)}}
\]

Table \ref{delta3} shows the obtained values whit the $\Delta_{Statistical}$ rounded to integers.

\begin{table*}
 \centering
 \begin{minipage}{170mm}
  \caption{Satellite Mean Monthly Percentage 2007.}
   \label{Mean07}
  \begin{tabular}{@{}lllllccccccccccc@{}}
  \hline
           &\multicolumn{2}{c}{Paranal} & \multicolumn{2}{c}{La Silla}& \multicolumn{2}{c}{La Palma} & \multicolumn{2}{c}{Mt.Graham} & \multicolumn{2}{c}{Tolonchar}\\
           & Clear& Stable & Clear& Stable& Clear& Stable& Clear& Stable& Clear& Stable\\

 \hline
 January     &73 &55 &72 &72 &38 &38 &48 &48 &54 &42\\
 February  & 90 &61 & 90 & 88 &61 &61 & 45 & 45 & 72 &62 \\
 March     &   86 & 65 & 75 & 75  &52 & 52 & 51 & 51 & 81 & 63\\
 April     &  74 & 58 & 64 &58 & 75 & 75 & 61 & 61 &94 & 64\\
 May       &  91 & 72 & 59 & 59 & 86 & 86 & 59 & 59 & 87 & 87\\
 June      &  60 & 55 & 33 & 33 & 94 & 88 & 72 & 72 & 69 & 69\\
 July      & 89 & 82 & 56 & 56 &93 & 86 & 14 & 14 & 82 & 82\\
 August    &  87 & 76 & 65 & 65 & 93 & 93 & 41 & 41 & 89 & 89\\
 September & 94 & 78 & 73 & 73 & 80 & 80 & 44 & 44 & 100 & 92\\
 October   &  100 & 91 & 93 & 90 & 80 & 80 & 72 & 69 & 93 & 84\\
 November  & 98 & 80 & 85 & 85 & 48 & 48 & 53 & 52 & 93 & 79  \\
 December  &  80 & 55 & 78 & 78 & 84 & 80 & 72 & 71 & 91 & 76\\
\hline
 Mean      & 85 & 69 & 70 & 69 & 74 & 72 & 53 & 52 & 84 & 74\\
 \hline

 Clear-Stable Mean &\multicolumn{2}{c}{77} &\multicolumn{2}{c}{70}   &\multicolumn{2}{c}{73} &\multicolumn{2}{c}{52} &\multicolumn{2}{c}{79} \\
 \hline

\end{tabular}
\end{minipage}
\end{table*}

\begin{table*}
 \centering
 \begin{minipage}{170mm}
  \caption{Satellite Mean Monthly Percentage 2008.}
   \label{Mean08}
  \begin{tabular}{@{}lllllccccccccccc@{}}
  \hline
           &\multicolumn{2}{c}{Paranal} & \multicolumn{2}{c}{La Silla}& \multicolumn{2}{c}{La Palma} & \multicolumn{2}{c}{Mt.Graham} & \multicolumn{2}{c}{Tolonchar}\\
           & Clear& Stable & Clear& Stable& Clear& Stable& Clear& Stable& Clear& Stable\\

 \hline
 January     &67 &57 & 100 & 68 & 52 & 52 & 49 & 49 & 52 & 52 \\
 February  & 95 & 73 & 96 & 74 & 68 & 68 & 60 & 60 & 85 & 81 \\
 March     & 100 & 63 & 93 & 74  &52 & 52 &86 & 80 & 99 & 74 \\
 April     &  94 & 61 & 82 & 82&77 & 77&93 & 78 & 100 & 80 \\
 May       & 90 & 72 & 71 & 71 & 87 & 82 & 75 & 70 & 91 & 77 \\
 June      &  78 & 65 &64 & 64 & 90 & 90 & 74 & 66 & 72 & 72 \\
 July      & 97 & 60 & 74 & 65 & 95 & 84 & 20 & 20 & 99 &88\\
 August    & 99 & 80 & 72 & 72 & 90 & 90 & 24 & 24 &95 & 91  \\
 September & 94 & 76 & 56 & 56 & 71 &71&74 & 74 & 100  & 97\\
 October   & 97 & 85 & 80 & 77 & 82 & 82 & 86 & 80 & 100 & 89 \\
 November  & 96 & 78 & 97 & 88 & 37 & 37 & 71 & 71 & 100 &83  \\
 December  &  88 & 75 & 100 & 85 &45 & 45 & 62 & 62 & 70 & 70 \\
\hline
 Mean      & 91 & 70 & 82 & 73 & 71 & 69 & 65 & 61 &89 & 80 \\
 \hline

 Clear-Stable Mean &\multicolumn{2}{c}{81} &\multicolumn{2}{c}{78}   &\multicolumn{2}{c}{70} &\multicolumn{2}{c}{63} &\multicolumn{2}{c}{84} \\
 \hline

\end{tabular}
\end{minipage}
\end{table*}

\begin{table}
 \centering
 \begin{minipage}{80mm}
  \caption{Mathematical and statistical uncertainties of the model in 2008 at Paranal and La Palma.}
   \label{delta3}
  \begin{tabular}{@{}lccc@{}}
  \hline

 Site                 & $\Delta_{Total}$  & $N(G;S)$ & $\Delta_{Statistical}$  \\

 \hline
 Paranal              & $7.3\%$           & 1050      &  $2.0\%$               \\
 La Palma (Temporal)     & $9.1\%$           & 1020     &   $3.0\%$               \\
 La Palma (Daily)      & $13.1\%$          & 340      &   $9.0\%$              \\
 \hline

\end{tabular}
\end{minipage}
\end{table}

\section{Conclusion}

In this paper we have presented a new homogeneous method in order to obtain the amount of available time fraction. The data are extracted from GOES12 satellite imager on five very important and different astronomical sites in order to get comparable statistics. Satellite data are compared with ground based data.\\
In this analysis a wider spatial field is used in order to reduce the spatial noise: each value is the mean of $1^\circ\times1^\circ$ matrix. The cloud coverage is obtained using GOES12 B4 and B6 bands independently. Using the correlation of three bands (B3, B4 and B6) we have computed an atmospheric correlation function as a further selection of the clear nights, and we have introduced the new concept of stable night. Temporal data are used for the years 2007 and 2008. We have shown that the derived atmospheric correlation function is correlated with the quality of the night in terms of FWHM (possibly also with wind and humidity).
An example of clear/stable nights is given in Fig. \ref{p} that shows the monthly distribution of 2008 nights at Paranal. The black bars represent the monthly percentage  of clear nights, while the gray bars the percentages of stable nights. We can assume that stable nights could be the best approximation to the photometric nights.
We obtained that the amount of stable nights is considerably lower than the clear nights in all the five analysed sites.
In view of a better tuning of the stable nights as a function of the ground based parameters, we can adopt as a best approximation of the "clear nights" of the ground based log the satellite clear night percentages.\\
The mean of the 2007-2008 give a percentage of clear time of 88\% at Paranal, 76\% at La Silla, 72.5\% at La Palma, 59\% at Mt. Graham and 86.5\% at Tolonchar. These percentage differences are higher than the statistical errors (Table \ref{delta3}).\\
Tolonchar and Paranal (Tables \ref{Mean07} and \ref{Mean08}) show the largest number of clear nights but Tolonchar shows the largest number of stable nights, while La Palma shows that if a night is clear is also almost stable.
Tolonchar appears the best site as concerning the stable nights while Paranal is the best for the clear nights (see also Figures \ref{s} and \ref{s4}).\\
The procedure adopted in this paper gives different
percentages of satellite clear nights when compared with those of Erasmus \&
van Rooyen (\cite{erasmus06}). In fact we found 88\% of clear nights at
Paranal to compare with the Erasmus's percentage of 85\%, instead, at La
Palma we found the 72.5\% to compare with the 83.7\% of Erasmus.
As already explained in the text the two methods differ mainly because (1)
we use the direct brightness values of the satellite while Erasmus \&
van Rooyen (\cite{erasmus06}) converted them into temperatures and
intepreted the absolute values of the temperatures in terms of height of
the infrared emission, using a temperature-height sounding, and then of
cloud coverage, and (2) they used a much smaller matrix. The authors
pointed out that this technique has some limitations due to a number of
effects, for example anomalous trends in temperatures during the night or
for some types of clouds (monsoon clouds are an example). While their
approach is certainly valid in terms of general physical interpretation,
we found more direct and more reliable to work directly in terms of
brigthness relative time fluctuations. A deep analysis should be done
comparing the results night by night but this is out of the scope of the
present paper.\\
On the other hand we should take into account possible biases due to our time sampling, because we are measuring the second part of the night only. Our limit was set by the time availability of the Paranal ground log. In particular, phenomena that occurred during the first part of the night were not analyzed in this paper.\\In addition we note that some low level phenomena could be missing (for example local dust clouds, fog...). A further possible bias is due to the satellite spatial resolution, mainly in sites with abrupt topography.\\
The use of higher resolution satellites, for example
the MERIS spectrograph (on board of Envisat) with a spatial resolution of
about 1 km, in principle should be better in these cases. Some authors
obtained interesting results. For example Kurlandczyk and Sarazin (\cite{kurlandczyk})
used MERIS at La Silla and Paranal to get cloud coverage and precipitable
water vapour and discussed the horography effects. However, in spite of
its high spatial resolution MERIS presents some disadvantages compared to
GOES. First the temporal coverage is much lower, second, MERIS is working
in daytime and it does not give data during the night. As a consequence it
can be used as a complement of GOES data to investigate the effects of
spatial resolution, but its generalized use should be carefully validated
site by site.\\
Finally, it is interesting to note the percentage differences between the two years, particularly at Mt.Graham, in 2008, the clear and stable nights percentage was considerably higher. We note minor differences also in the other sites. These could be a result of the El $Ni\widetilde{n}o$ phenomenon and its consequences at different sites.\\
It is possible to reduce the uncertainty of this methodology using  all the available GOES12 IR bands and refining the tuning of the model.
We found that using the correlation function from IR satellite data, it is also possible to observe
several atmospheric phenomena (i.e. strong winds, damp winds, warm winds, fogs, humidity, dust etc).\\
A second paper on this correlation analysis is in progress to study of the cyclical fluctuations of this function to test the possibility to have a sort of now-casting seeing.\\ The possible synergy of this model with  seeing forecast models may predict the atmospheric changes in the short and long time-scale, allowing the atmospheric conditions for science cases in order to have the best scientific results.

\subsection{ACKNOWLEDGMENTS}

The authors acknowledge dr.Vincenzo Testa, from National Institute
for Astrophysics, Roland Gredel, from Heidelberg Max Planck
Institute for Astronomy, Antonia M. Varela, from Instituto de Astrofisica de Canarias, Antonio Magazzu from TNG and the former TNG
Director Ernesto Oliva for the collaboration.\\
This activity is supported by the European Community (Framework
Programme 7, Preparing for the construction of the European
Extremely Large Telescope, Grant Agreement Number 211257) and by Strategic University of Padova funding by title "QUANTUM FUTURE".\\
Most of data of this paper are based on the CLASS (Comprehensive Large Array-data Stewardship System).\\ CLASS is an electronic library of NOAA environmental data.\\ This web site provides capabilities for finding and obtaining those data, particularly NOAA's Geostationary Operational Environmental Satellite data.\\
Finally we acknowledge the Liverpool Telescope website staff.

\subsection{List of Acronyms}

\begin{itemize}
	
	\item GOES: Geostationary Operational Environmental Satellite 
	\item MERIS: Medium Resolution Imaging Spectrometer
	\item CLASS: Comprehensive Large Array-data Stewardship System
	\item ESO: European Southern Observatory
	\item TNG: Telescopio Nazionale Galileo
	\item TMT: Thirty Meter Telescope
	\item INT: Isaac Newton Telescope
	\item ING: Isaac Newton Group of Telescopes
	\item DIMM: Differential Image Motion Monitor
	\item RoboDIMM: Robotic Differential Image Motion Monitor

\end{itemize}

\label{lastpage}

\end{document}